\documentclass[
   prd,
   amsfonts,
   amssymb,
   amsmath,
   preprint,
   eqsecnum,
   showpacs,
   byrevtex]{revtex4}
\usepackage{color}
\usepackage{graphicx}
\usepackage{subfigure}
\newcommand{\beq}{\begin{equation}}
\newcommand{\bea}{\begin{eqnarray}}
\newcommand{\ee}{\end{equation}}
\newcommand{\eea}{\end{eqnarray}}

\newcommand{\rd}{\mathrm{d}}               
\newcommand{\kperp}{k_{\perp}}             
\newcommand{\keta}{k_{\eta}}               
\newcommand{\Dslash}{D \kern -0.65em/ \ }  
\newcommand{\bra}[1]%
   {\ensuremath{\langle \, #1 \, |}}
\newcommand{\bigbra}[1]%
   {\ensuremath{\Bigl \langle \, #1 \, \Bigr |}}
\newcommand{\ket}[1]%
   {\ensuremath{| \, #1 \, \rangle}}
\newcommand{\bigket}[1]%
   {\ensuremath{\Bigl | \, #1 \, \Bigr \rangle}}
\newcommand{\braket}[2]%
   {\ensuremath{\langle \, #1 \, | \, #2 \, \rangle}}

\newcommand{\Expect}[1]
   {\ensuremath{\langle \, #1 \,  \rangle}}
\newcommand{\Comm}[2]
   {\ensuremath{[ \, #1, #2 \, ]}}
\newcommand{\AntiComm}[2]
   {\ensuremath{\{ \, #1, #2 \, \}}}
\newcommand{\Tr}[1]{\mathrm{Tr} [ \, #1 \, ]}      

\newcommand{\Exp}[1]{\exp [ \, #1 \, ]}            
\newcommand{\Ln}[1]{\ln [ \, #1 \, ]}              
\begin{document}
\title{Real time particle production in QED and QCD
from strong fields and the Back-Reaction problem:  \\
talk given by Fred Cooper at the KITP conference on nonequilibrium phenomena, February, 2008}
\author{Fred Cooper}
\email{cooper@santafe.edu}
\affiliation{Santa Fe Institute,
   Santa Fe, NM 87501}
\affiliation{National Science Foundation,
   4201 Wilson Blvd.,
   Arlington, VA 22230}
\author{John F. Dawson}
\email{john.dawson@unh.edu}
\affiliation{Department of Physics,
   University of New Hampshire,
   Durham, NH 03824}
\author{Bogdan Mihaila}
\email{bmihaila@lanl.gov}
\affiliation{Materials Science and Technology Division,
   Los Alamos National Laboratory,
   Los Alamos, NM 87545}
%
%
\begin{abstract}
We review the history of analytical approaches to particle production from external strong fields  in QED and QCD , and numerical studies of the  back reaction problem for the electric field in QED. We discuss the formulation of the backreaction problem for the chromoelectric field in QCD both in leading and next to leading order in flavor large-N QCD.
\end{abstract}
\maketitle
%
%
\tableofcontents
\newpage
%
%
\section{Introduction}
\label{s:intro}

Nonequilibrium quantum field theory became a topic of interest as a result of several crucial questions facing physics in the 1980's.  Before that time, people were mainly concerned with calculating cross sections (S-Matrix elements) for particles produced at accelerators.  In the early 80's the idea of the Inflationary universe was introduced by Guth, Linde and Starobinsky  \cite{inflation} and the mechanism for inflation was a scalar field (the inflaton) evolving in time  in a background (time dependent) gravitational field.  The initial calculations were done using a classical scalar field. It was natural for people to then try to understand how to solve time evolution problems in the quantum domain.  In another area of Cosmological importance, Emil Mottola looked at the problem of particle production of ``free'' (apart from Gravity) scalar mesons in De Sitter Space  as a possible mechanism for the solution of the cosmological constant problem. \cite{Mottola}.  These two problems were the basis for renewed interest in the problem of how one solves numerically initial value and backreaction problems in quantum field theory. At that time even the correct formalism for doing this was not well understood by particle physicists.  In trying to solve  this problem,  one has the choice of the Schrodinger picture (functional Schrodinger equation) or the Heisenberg picture (path integral or Schwinger-Dyson
equations). Even though, in the Heisenberg picture,  Schwinger and Keldysh  \cite{ctp} had introduced the correct formalism for treating this problem, the closed time path formalism (CTP),  at that time very few relativistic  field theory calculations had been attempted  and the question of how to do practical calculations which required renormalization had not been addressed. Also at that time it was not realized that if one used perturbation theory in the CTP  formalism, that the resulting perturbation series (beyond leading order) was secular \cite{secular}.

The first study of inflation which included quantum effects was In 1985, when Guth and Pi \cite{Pi} studied the problem of the quantum roll from the top of  an inverted harmonic oscillator in the free field limit to see to what extent classical ideas on inflation were modified by quantum effects.  The formalism used to study this problem was the functional Schrodinger equation which was the generalization of the  ordinary Schrodinger equation to quantum field theory.  This study however did not require renormalization.   The functional Schrodinger equation could be derived by the Dirac action principle.  This then allowed people to study time the quantum evolution process using gaussian trial wave functions (i.e. variational approximations).  This approximation to the field theory  was  related to solving the field theory in the Heisenberg picture  in a mean field  (Hartree approximation or leading order in large-N approximation).  The first conceptual problems that needed to be solved in  the scalar field theory in this approximation were how to disentangle the infiinities that arose from choosing (potentially unphysical) initial data  from those that were truly effects of renormalization. These issues and their solution was clarified in the work of Cooper and Mottola and Samiullah and Pi \cite{solution}. The initial value problem was solved in the gaussian (leading order in large-N) approximation by choosing initial states that corresponded to finite energy density and number density with respect to the vacuum of the
gaussian approximation.  The renormalization issues were first understood by doing a WKB analysis of an adiabatic expansion of the Green's functions. This approach was related to the method of adiabatic regulation used to study free fields in background gravitational fields \cite{Fulling}.   Later it was shown that one could study the renormalization group flow of the coupling constant with momentum and verify that a more standard renormalization which looked for the discrete momentum space regulated answer converging to the continuum renormalization being the simplest approach \cite{renorm2}.  Although the functional Schrodinger equation coupled with variational ansatzes led to
approximations which conserved probability and energy,unlike  what happens in the case of variational approximations in quantum mechanics, it was difficult to find ways to go beyond the Gaussian approximation in a systematic fashion.  On the other hand the CTP formalism has a nice path integral representation which allowed for ordinary perturbation expansions as well as $1/N$ expansions which were systematic.  The downside of the CTP formalism was that the latter two expansions directly applied to the path integral turned out to be secular and did not exactly conserve probabilities.  This serious defect would later be cured by looking at expansions based on the generating functional of two particle irreducible graphs.

It was in order to better understand back reaction as a solution of the cosmological constant problem that Cooper and   Mottola  first studied as a ``toy'' model, backreaction in the Electric Field case.  \cite{kluger1}. The degradation of the electric field as a result of pair production was the test case for the degradation of the cosmological constant as a result of pair production.   While this pair production problem was being
investigated, there was an independent compelling reason for studying this problem.  In order to understand particle production following relativistic heavy ion collisions people began investigating in the context of transport theory,  a flux tube model based on Schwinger pair production,  which converted this toy problem into one with experimental consequences \cite{flux}.  In light of the renewed interest in particle production from semi-classical gluonic fields we have recently undertaken a study of the quantum back reaction problem in  SU(3) QCD in 3+1 dimensions in the hope of seeing what one can learn about  jet production at RHIC and LHC and also  the initial gluon condensate state.

In this talk we would first like to review what is known analytically about pair production from strong
electric and chromoelectric fields, both static and time dependent.  Then we will concentrate on making the problem more realistic by solving simultaneously as an initial value problem the Dirac equation in an external electric field, and the update equation for the electric field as it degrades and oscillates due to expansion and particle production.
 We will first  review previous results on pair production and back reaction in QED mostly in 1+1 dimension, where the kinematics relevant for Heavy Ion collisions were used.   Next we will discuss current work on solving the back reaction and particle production problem ins SU(3) QCD in 3+1 dimensions.  Finally
 we will discuss how one would go about adding  interactions so that  the issue of thermalization of the fields during expansion can be  addressed.

%
%
\section{Pair production from a strong electric and chromoelectric fields}

In order to pop a pair of fermions (or bosons) out of the vacuum  using strong Electric Fields one  must supply an energy  $e E x$ in a Compton wave length $x \approx \hbar /mc$ which is at least twice the rest energy of the pair ( $2 m c^2$) . The  critical value of the electric field for this to happen is of order
\beq
 e E \hbar/mc = 2 mc^2
\ee
 or
 \beq
 E _{critical} \approx   2 m^2 c^3 / e \hbar .
 \ee
 This critical electric field is not yet attainable using lasers, but the analogue process of producing chromoelectric fields by colliding heavy ions does lead to initial energy densities that are above the critical value.  The dimensionless variable relevant for this process is thus
 \beq
\frac{E}{E_{critical}} = \frac{ E} {2 e \hbar m^2 c^3}
 \ee
  In order to see how this  variable arises, one can make the following simple tunneling  picture  \cite{itz} of the non-perturbative process of pair production.
One imagines that one has an electron bound in a potential well of order  $|V_0| \approx 2m c^2$ and
one then applies a constant electric field which leads to a one dimensional extra potential of  $eEx$
to the (say square well) potential of depth $2 m c^2$.  On then finds that the ionization probability is proportional to the WKB barrier penetration factor:
\beq
\exp \left[ -2 \int_0^{V_0/eE} dx \left(2m (V_0 - |e E| x)^{1/2} \right) \right] = e^{-(8m^2 c^3/ 3e \hbar E)}.
\ee
In what follows we will set  $\hbar=1; c=1$.
We see that when $E > E_{critical}$ there is no exponential suppression of pair production.
A more careful calculation discussed below involves determining the imaginary part of the vacuum persistence function in the presence of the external field.

%
%
\subsection{Constant electric and chromoelectric field results}

In his classic paper in 1951 Schwinger derived the following one-loop non-perturbative formula
\bea
\frac{dW}{d^4x}=\frac{e^2E^2}{4\pi^3} ~\sum_{n=1}^{\infty}  ~\frac{1}{n^2}~e^{-\frac{n\pi m^2}{|eE|}}
\eea
for the probability of $e^+e^-$ pair production per unit time per unit volume from a constant electric field E via vacuum polarization \cite{schw} by using a  proper time method. In case of charged
scalar field theory the corresponding result is given by
\bea
\frac{dW}{d^4x}=\frac{e^2E^2}{8\pi^3} ~\sum_{n=1}^{\infty} ~\frac{(-1)^{n+1}}{n^2}~e^{-\frac{n\pi m^2}{|eE|}}.
\eea

The result of Schwinger was extended to QCD by Claudon, Yildiz and Cox \cite{yildiz1}.  However the $p_T$ distribution of the $e^+$ (or $e^-$) production, $dW / ( d^4x d^2p_T )$, could not be obtained by using proper time method of Schwinger.  A WKB approximate method was used for this purpose by Casher et. al. \cite{Casher},  but an exact method to do this problem (of determining the transverse distribution of pairs) was not found until recently \cite{gouranga}.  For QED the WKB analysis gave the correct answer which depended only on the
energy density of the electric field.  However for QCD, the WKB answer was similar to the exact answer for QED but incorrect  for QCD in that
the exact answer depended on both Casimir invariants of $SU(3)$.
In the case of fermions in QED one finds for the transverse distribution of fermion pairs:
\bea
\frac{dW}{d^4xd^2p_T}=-\frac{|eE|}{4\pi^3} {\rm Log}[1-e^{-\pi \frac{p_T^2+m^2}{|eE|}}].
\label{dsf}
\eea
The corresponding result for the charged scalar production is given by
\bea
\frac{dW}{d^4xd^2p_T}=\frac{|eE|}{8\pi^3} {\rm Log}[1+e^{-\pi \frac{p_T^2+m^2}{|eE|}}].
\label{ds}
\eea

In QCD the transverse distribution instead
depends on two independent Casimir invariants of SU(3): $C_1=[ \, E^aE^a \, ]$ and
$C_2=[ \, d_{abc}E^aE^bE^c \, ]^2$, where $E^a$ is the constant chromo-electric field with color
index a=1,2,..8 \cite{gouranga}.
Nayak obtained the following formula for the number of non-perturbative quarks
(antiquarks) produced per unit time, per unit volume and per unit transverse
momentum from a given constant chromo-electric field $E^a$
\bea
\frac{dN_{q,\bar q}}{dt d^3x d^2p_T}~
=~-\frac{1}{4\pi^3} ~~ \sum_{j=1}^3 ~
~|g\lambda_j|~{\rm ln}[1~-~e^{-\frac{ \pi (p_T^2+m^2)}{|g\lambda_j|}}]\,,
\label{nayak1}
\eea
where $m$ is the mass of the quark.
This result is gauge invariant because it depends on the
following gauge invariant eigenvalues
\bea
&&~\lambda_1~=~\sqrt{\frac{C_1}{3}}~{\cos}\theta\,,
\nonumber \\
&&~\lambda_2~=~\sqrt{\frac{C_1}{3}}~{\cos}~ (2\pi/3-\theta)\,,
\nonumber \\
&&~\lambda_3~=~\sqrt{\frac{C_1}{3}}~{\cos}~ (2\pi/3+\theta)\,,
\label{lm}
\eea
where $\theta$ is given by
\bea
\cos^23\theta~=3C_2/C_1^3\,.
\label{theta}
\eea

The integration over $p_T$ in eq. (\ref{nayak1})
reproduces Schwinger's proper time result, extended to QCD, for total production rate $dN/d^4x$
\cite{yildiz1}.
The exact result in eq. (\ref{nayak1}) can be contrasted with the following
formula obtained by the WKB tunneling method \cite{Casher}
\bea
\frac{dN_{q,\bar q}}{dt d^3x d^2p_T}~=~\frac{-|gE|}{4\pi^3} ~
{\rm ln}[1~-~e^{-\frac{\pi (p_T^2+m^2)}{|gE|}}]\,,
\label{7}
\eea
which does not reproduce the correct result for
the $p_T$ distribution of the quark (antiquark) production rate from a
constant chromo-electric field $E^a$.
For soft gluon production Nayak and van Nieuwenhuizen \cite{gouranga} found in the Feynman T'Hooft gauge \cite{thooft}
\bea
\frac{dN_{gg}}{dt d^3x d^2p_T}~
=~\frac{1}{4\pi^3} ~~ \sum_{j=1}^3 ~
~|g\lambda_j|~{\rm ln}[1~+~e^{-\frac{ \pi p_T^2}{|g\lambda_j|}}].
\eea
This was shown to be independent of the gauge fixing choice  by Cooper and Nayak  \cite{independent}.

%
%
\subsection{Schwinger pair production in $SU(3)$ gauge
theory}

In the background field method of QCD the gauge field is the sum of a classical
background field and  the quantum gluon field:
\bea
A_\mu^a ~\rightarrow ~ A_\mu^a~+~Q_\mu^a
\label{aq}
\eea
where in the right hand side $A_\mu^a$ is the classical background field
and $Q_\mu^a$ is the quantum gluon field. The gauge field Lagrangian density is
given
by
\bea
{\cal L}_{\rm gauge}
   =
   -\frac{1}{4} F_{\mu \nu}^a[A+Q] \, F^{a;\mu \nu}[A+Q].
\label{total}
\eea
The background gauge fixing is given by  by \cite{thooft}
\bea
   D_\mu[A] \, Q^{\mu a}
   = 0 \>,
\label{gfix}
\eea
where the covariant derivative is defined by
\bea
D_\mu^{ab}[A]~=~\delta^{ab} \partial_\mu~+~gf^{abc}A_\mu^c.
\eea
The gauge fixing Lagrangian density is
\bea
   {\cal L}_{\rm gf}
   =
   -\frac{1}{2\alpha} \,
   \bigl [ \, D_\mu[A] \, Q^{\mu a} \, \bigr ]^2
\label{gf}
\eea
where $\alpha$ is any arbitrary gauge parameter, and the corresponding ghost contribution is given by
\beq
{\cal{L}}_{\text{ghost}}
   =
   \overline{\chi}^a D_{\mu}^{ab}[A] \,
   D^{\mu ,bc}[A+Q]\chi^c
   =
   \overline{\chi}^a ~ K^{ab}[A,Q] ~ \chi^b \>.
\label{ghos1}
\ee
Now adding Eqs.~(\ref{total}), (\ref{gf}), and (\ref{ghos1}), we get the Langrangian density for gluons interacting with a classical background field:
\begin{equation}\label{lgluon}
   \mathcal{L}_{\text{gluon}}
   =
   -
   \frac{1}{4} \, F_{\mu \nu}^a[A+Q] \, F^{a;\mu \nu}[A+Q]
   -
   \frac{1}{2\alpha} \,
   \bigl [ \,
      D_\mu[A] \, Q^{a;\mu} \,
   \bigr ]^2
   -
   \overline{\chi}^a \, K^{ab}[A,Q] \, \chi^b.
\end{equation}
To discuss gluon pair production at the one-loop level on considers just the part of this
Lagrangian which is quadratic in quantum fields.  This quadratic Lagrangian is invariant
under a restricted class of gauge transformations.
The quadratic Lagrangian  for
a pair of gluon interacting with background field $A_\mu^a$ is given by
\bea
   {\cal L}_{\text{gg}}
   =
   \frac{1}{2} \, Q^{\mu a} \, M^{ab}_{\mu \nu}[A] \, Q^{\nu b}
\label{fulll}
\eea
where
\bea
   M^{ab}_{\mu \nu}[A]
   =
   \eta_{\mu \nu} [D_\rho(A)D^\rho(A)]^{ab}
   -
   2gf^{abc}F_{\mu \nu}^c
   +
   \Bigl ( \, \frac{1}{\alpha} - 1 \, \Bigr ) \,
   [D_\mu(A) \, D_\nu(A)]^{ab}
\label{mab}
\eea
with $\eta_{\mu \nu}~=~(-1,+1,+1,+1)$.

For our purpose we write
\bea
   M^{ab}_{\mu \nu}[A]
   =
   M^{ab}_{\mu \nu;\alpha=1}[A]
   +
   \alpha' \, [D_\mu(A) \, D_\nu(A)]^{ab}
\label{mab1}
\eea
where $\alpha^\prime~=~(\frac{1}{\alpha}-1)$. The matrix elements for the
gauge parameter $\alpha$=1 is given by
\bea
   M^{ab}_{\mu \nu;_\alpha=1}[A]
   =
   \eta_{\mu \nu} \, [D_\rho(A)D^\rho(A)]^{ab}
   -
   2gf^{abc} \, F_{\mu \nu}^c
\label{mabalpha1}
\eea
which was studied in \cite{gouranga}.
In this approximation,  the ghost Lagrangian density is given by
\beq
   \mathcal{L}_{\text{ghost}}
   =
   \overline{\chi}^a D_{\mu}^{ab}[A]
   D^{\mu ,bc}[A]\chi^c~=~\overline{\chi}^a~K^{ab}[A]~\chi^b
\label{ghos2}
\ee
The vacuum-to-vacuum transition amplitude in pure gauge theory
in the presence of a background field $A_\mu^a$ is given by:
\bea
   {}_{+} \langle \, 0 \, | \, 0 \, \rangle^A_{-}
   =
   \int [dQ] \,
   [d\chi] \, [d\bar{\chi}] \,
   e^{i{( S + S_{\text{gf}} + S_{\text{ghost}}) } } \>.
\eea
For the gluon pair part this can be written by
\bea
   {}_{+} \langle \, 0 \, | \, 0 \, \rangle^A_{-}
   =
   \frac{Z[A]}{Z[0]}
   =
   \frac{  \int [dQ]~e^{i\int d^4x \,
   Q^{\mu a} \,
   M_{\mu \nu}^{ab}[A] \,
   Q^{\nu b}} }
   { \int [dQ] \, e^{i\int  d^4x \,
   Q^{\mu a} \, M_{\mu \nu}^{ab}[0] \, Q^{\nu b}}}
   =
   e^{iS^{(1)}_{\text{eff}}}
\label{vac1}
\eea
where $S^{(1)}_{\text{eff}}$ is the one-loop effective action. The non-perturbative real gluon production is related to the
imaginary part of the effective action $S^{(1)}_{\text{eff}}$ which is physically due to the
instability of the QCD vacuum in the presence of the background field.
The above equation can be written as
\bea
   {}_{+} \langle \, 0 \, | \, 0 \, \rangle^A_{-}
   =
   \frac{Z[A]}{Z[0]}
   =
   \frac{{\rm Det^{-1/2}}M_{\mu \nu}^{ab}[A]}
        {{\rm Det^{-1/2}}M_{\mu \nu}^{ab}[0]}
   =
   e^{iS^{(1)}_{\text{eff}}}
\label{vac2}
\eea
which gives
\bea
   S^{(1)}_{\text{eff}}
   =
   -i {\rm Ln}
   \frac{(\rm Det [M_{\mu \nu}^{ab}[A]])^{-1/2}}
      {(\rm Det [M_{\mu \nu}^{ab}[A]])^{-1/2}}
   =
   \frac{i}{2} \,
   \Tr{ {\rm Ln} M_{\mu \nu}^{ab}[A]
        - {\rm Ln} M_{\mu \nu}^{ab}[0] } \>.
\label{vac3}
\eea
The trace contains an integration over $d^4x$ and
a sum over color and Lorentz indices.
To the above action, we need to add the ghost action.
The ghost action is gauge independent and eliminates the unphysical gluon degrees of
freedom.  The one-loop effective action for the ghost in the background field $A_\mu^a$
is given by
\bea
S^{(1)}_{\text{ghost}}
   =
   -i {\rm Ln}(Det~K)
   =
   -i \, \mathrm{Tr} \int_0^\infty~\frac{ds}{s} \,
   \Bigl [ \,
      e^{is~[K[0]+i\epsilon] } - e^{is~[K[A]+i\epsilon]} \,
   \Bigr ]
\eea
where $K^{ab}[A]$ is given by (\ref{ghos2}).
Since the total action is the sum of the
gluon and ghost actions, the gauge parameter dependent part proportional
to $( \, 1 / \alpha - 1 \,)$ can
be evaluated as an addition to the $\alpha=1$.

The non-perturbative gluon pair production per unit volume per unit time is
related to the imaginary part of this effective action via
\bea
   \frac{dN}{dtd^3x}
   \equiv
   {\rm Im} \, {\cal L}_{\text{eff}}
   =
   \frac{{\rm Im} \, S^{(1)}_{\text{eff}}}{d^4x}.
\label{nongluon}
\eea
This is the general formulation of
Schwinger mechanism in pure gauge theory where
$M^{ab}_{\mu \nu}[A]$ is given by eq. (\ref{mab1}) and
${M^{ab }_{\mu \nu,}}_{\alpha=1}[A]$ is given by eq. (\ref{mabalpha1}). In \cite{gouranga} this
expression was evaluated for $\alpha=1$ and the final expression for the number of non-perturbative
gluon (pair) production per unit time per unit volume and per unit transverse
momentum from constant chromo-electric field $E^a$ is given by \cite{gouranga}

\bea
\frac{dN_{g,g}}{dt d^3x d^2p_T}~
=~\frac{1}{4\pi^3} ~~ \sum_{j=1}^3 ~
~|g\lambda_j|~{\rm Ln}[1~+~e^{-\frac{ \pi p_T^2}{|g\lambda_j|}}].
\eea
After this calculation was done in $\alpha=1$ gauge, Cooper and Nayak showed by explicit
evaluation of the extra term proportion to $\alpha-1$ that the result for the particle production rate
was independent of the gauge fixing parameter $\alpha$
\cite{independent}.

Recently there has also been some progress to extending this result to time dependent fields.  Using a formal operator shift theorem  \cite{shift},  Nayak and  Cooper were able to show:

\bea
\frac{dW}{d^4xd^2p_T}=\frac{|eE(t)|}{8\pi^3} {\rm Log}[1+e^{-\pi \frac{p_T^2+m^2}{|eE(t)|}}].
\label{dw}
\eea
For Fermion pair production we get instead:
\bea
\frac{dW}{d^4xd^2p_T}=-\frac{|eE(t)|}{4\pi^3} {\rm Log}[1-e^{-\pi \frac{p_T^2+m^2}{|eE(t)|}}].
\label{dwf}
\eea
These results came from evaluating the one loop Action.  For the Boson case we obtained
\bea
S^{(1)}_B=
\frac{i}{16\pi^3}
\int_0^\infty \frac{ds}{s} \int d^4x \int d^2p_T
e^{is(p_T^2+m^2+i\epsilon)}
[ \frac{1}{ s}- \frac{eE(t)}{{\rm sinh}(seE(t))}].
\label{12fff}
\eea
wheras in the fermion case we obtained instead:
\bea
S^{(1)}=
\frac{i}{8 \pi^3}
\int_0^\infty \frac{ds}{s} \int d^4x \int d^2p_T
e^{is(p_T^2+m^2+i\epsilon)}
[eE(t) ~{\rm coth}(seE(t))~-~\frac{1}{s}].
\label{12fg}
\eea

The way the above results were obtained was to start with the effective action
for scalar field theory  where
\[  M[A]=({\hat p}-eA)^2-m^2;   ~~~{\hat p}_\mu = i\frac{\partial}{\partial x^\mu}  \]
The Action is
\[ S^{(1)}=i{\rm Tr ln} [({\hat p}-eA)^2 -m^2] -i {\rm Tr ln} [\hat{p}^2  -m^2] \]
\[ =-i\int_0^\infty \frac{ds}{s} \int d^4x <x|[
e^{-is[(\hat{p}-eA)^2-m^2-i\epsilon]} -e^{-is(\hat{p}^2-m^2-i\epsilon)}]|x> \]
Choosing Axial gauge $A_3=0$  and  $A_0 =- E(t)z$
\begin{multline}
S^{(1)}
   =-i \int_0^\infty \frac{ds}{s}
   \int_{-\infty}^{+\infty} dt \,
   \langle \, t \, |
   \int_{-\infty}^{+\infty} dx \,
   \langle \, x \, |
   \int_{-\infty}^{+\infty} dy \,
   \langle \, y \, |
   \int_{-\infty}^{+\infty} dz
   \\
   \langle \, z \, | \,
   \Bigl \{ \,
    e^{-is \, [ \, ( \,
      \hat{p}_0+eE(t) z)^2
      -
      \hat{p}_z^2-\hat{p}_T^2
      -
      m^2-i\epsilon]}
      -
      e^{-is(\hat{p}^2-m^2-i\epsilon \, ) } \,
   \Bigr \} \,
    |\, z \, \rangle \, | \, y \, \rangle \,
    | \, x \rangle \, | \, t \, \rangle
 \end{multline}
Inserting complete set of $|p_T \, \rangle$ states
($\int d^2p_T \, |p_T\rangle \, \langle p_T | = 1$) and using
$\langle q|p \rangle =\frac{1}{\sqrt{2\pi}}e^{iqp}$ we obtain
\begin{multline}
   S^{(1)}
   =
   \frac{-i }{(2\pi)^2} \,
   \int_0^\infty \frac{ds}{s} \int d^2x_T \int d^2p_T \;
   e^{ is \, ( \, p_T^2+m^2+i\epsilon \, ) }
   \\ \times
   \Bigl \{ \,
      \int_{-\infty}^{+\infty} dt \,
      \langle \, t \,
      | \,
      \int_{-\infty}^{+\infty} dz \,
      \langle \, z \, |
       e^{ - is \, [ \,
          (\,
          - i\frac{d}{dt} + eE(t) z \,
          )^2
          -
          \hat{p}_z^2 \, ] } |\, z\, \rangle \,
          |\, t \, \rangle
          -
          \int dt \int dz \, \frac{1}{4\pi s}
   \Bigr \} \>.
\end{multline}
This expression contains the noncommuting quantities $E(t)$ and $d/dt$.  To evaluate these terms we derived a new shift theorem for operators:
\begin{multline}
   \int_{-\infty}^{+\infty} dx \,
   \langle \, x \, | \,
   e^{-[(a(y)x+h\frac{d}{dy})^2+b(\frac{d}{dx})+ c(y)]} \,
   | \, x \, \rangle \, f(y)
   \\
   = \int_{-\infty}^{+\infty} dx \,
   \Big \langle \, x \,
   -
   \frac{h}{a(y)}\frac{d}{dy} \,
   \Big | \,
   e^{-[a^2(y)x^2+b(\frac{d}{dx})+c(y)]} \,
   \Big | \,
   x
   -
   \frac{h}{a(y)} \, \frac{d}{dy} \,
   \Big \rangle \, f(y) \>.
\end{multline}
This  Shift Theorem also implies the interesting result.
 \begin{equation}
   \int_{-\infty}^{+\infty} dx \,
   e^{-(f(y)x+\frac{d}{dy})^2} \, g(y) \,
   =
   \int_{-\infty}^{+\infty} dx \, e^{-f^2(y)x^2}g(y)
   =
   \sqrt{\pi} \; \frac{g(y)}{f(y)}
\end{equation}
The results we found for the time dependent field have the remarkable feature that they are equivalent to Schwinger's original expressions for the effective action with the substitution $E \rightarrow E(t)$.  That is the adiabatic approximation
appears to give the exact result for the action.  Although this is initially surprising, it is not without
precedent.  A related result for the one loop effective action was found recently by Fried and Woodard  \cite{Woodard}, using Fradkin's formulation of the path integral,  for the case of an electric field pointing in the $z$ direction which arbitrarily depends on the light cone time coordinate $x^+ = (x^0+x^3)$.
Explicitly they found that the action integrated over momentum was also equivalent to the adiabatic result in the variable $x^+$, namely for the fermion action they obtained:
\begin{equation}
   \Gamma_1[A]
   =
   -i L[A]
   =
   \frac{1}{8 \pi^2} \,
   \int d^4x \int_0^{\infty}
   {ds \over s^3} \, e^{-i s m^2} \,
   \Bigl \{ \,
      e \, E(x^+) \, s \,
      \coth \bigl ( \, e E(x^+)s \, \bigr ) - 1 \,
   \Bigr \} \>.
\end{equation}

%
%
\section{Production and time evolution of a quark-antiquark plasma}

Our model for the production of the quark-gluon plasma begins with the
creation of a flux tube containing a strong color electric field. If
the energy density of the chromoelectric field gets high enough as discussed earlier,
 the quark-anti quark pairs (or gluon pairs)  can be popped out of the vacuum by tunneling.  For simplicity, first we
discuss pair production (such as electron-positron pairs) from an
abelian Electric Field and the subsequent quantum back-reaction on the
Electric Field. We will give extensive results for $1+1$ dimensions \cite{kluger2}.  In the next section we will discuss
extending this work to QCD in 3+1 dimension.  Our initial value problem will be that the Electric
(or Chromelectric) field is given at an initial (proper) time, and the fermions initially are in the vacuum state (no particles initially present).    Having a semi-classical Electric field interacting with fully quantum fermions is the first term in a large-N expansion of N-flavored QED.
\cite{largeN}. We assume as a reasonable first approximation, guided by hydrodynamical considerations,  that the kinematics of ultrarelativistic high energy collisions results in a boost invariant dynamics  \cite{boost} in the longitudinal
($z$) direction (here $z$ corresponds to the axis of the initial
collision) so that all expectation values are functions of the proper
time $\tau = \sqrt{t^2-z^2}$.We introduce the light cone variables
$\tau$ and $\eta$, which will be identified later with fluid proper
time and rapidity . These coordinates are defined in terms of the
ordinary lab-frame Minkowski time $t$ and coordinate along the beam
direction $z$ by
\begin{equation}
       z= \tau \sinh \eta  \quad ,\quad t= \tau \cosh \eta \,.
\label{boost_tz.tau.eta}
\end{equation}
 The Minkowski line element in these coordinates has the form
\begin{equation}
{ds^2} = {- d \tau^2 + dx^2   +dy^2 +{\tau}^2 {d \eta}^2 }\,.
\label{boost_line_element}
\end{equation}
Hence the metric tensor is given by
\begin{equation}
 g_{\mu \nu} = {\rm diag} (-1, 1, 1, \tau^2).
\end{equation}

The QED  action in curvilinear coordinates is:
\begin{eqnarray}
S &&= \int d^{d + 1}x \, ({\rm{det}}\, V) [ -{i \over 2}
\bar {\Psi} \tilde{\gamma}^{\mu}
\nabla_{\mu} \Psi+ {\frac{i}{2}} (\nabla^{\dag}_{\mu}\bar {\Psi} )
\tilde{\gamma}^{\mu} \Psi  \nonumber \\
&& -i m \bar {\Psi}\Psi- {1 \over 4}F_{\mu \nu}F^{\mu \nu} ],
\label{boost_Sf}
\end{eqnarray}
where
\begin {equation}
\nabla_{\mu} \Psi \equiv (\partial_{\mu} + \Gamma_{\mu} -ieA_{\mu})
\Psi
\end{equation}

Varying the action leads to the Heisenberg field
equation:
\begin{equation}
\left( \tilde{\gamma}^{\mu}\nabla_{\mu} + m \right) \Psi=0\,,
\end{equation}

\begin{equation}
\left[ \gamma ^0 \left(\partial_\tau+{1 \over 2\tau}\right)
+{\bf \gamma}_\perp\cdot \partial_\perp
+ {\gamma^3 \over \tau}(\partial_\eta -ieA_\eta)+ m \right] \Psi =0\,,
\label{boost_Dirac}
\end{equation}
 and the Maxwell equation:
$E=E_z(\tau)= - \dot{A}_{\eta}(\tau)$
\begin{equation}
{1 \over \tau}  {dE(\tau) \over d\tau} =  {e \over 2} \left \langle
\left[ \bar{\Psi}, \tilde {\gamma}^{\eta} \Psi \right] \right \rangle
={e \over 2 \tau} \left \langle \left[ \Psi^{\dagger}, \gamma^0
\gamma^3 \Psi \right] \right \rangle .
\label{boost_MaxD2}
\end{equation}

%
\vspace{.2cm}
%
%
We expand the fermion field in terms of Fourier modes at fixed proper
time: $\tau$,
\begin{eqnarray}
\Psi (x) &=& \int [d{\bf k}] \sum_{s}[b_{s}({\bf k})
\psi^{+}_{{\bf k}s}(\tau)
 e^{i k \eta} e^{ i {\bf{p}} \cdot {\bf x}} \nonumber \\
&&+d_{s}^{\dagger}({\bf{-k}}) \psi^{-}_{{\bf{-k}}s}(\tau)
e^{-i k \eta} e^{ - i {\bf{p}} \cdot {\bf x}}  ].
\label{boost_fieldD}
\end{eqnarray}
The $\psi^{\pm}_{{\bf k}s}$ then obey
\begin{eqnarray}
 \left[\gamma^{0} \left({d\over d \tau}+{1 \over 2\tau}\right)
+ i \gamma_{\bf{{\perp}}}\cdot{\bf{k_{\perp}}}
+i {\gamma^{3}} \pi_{\eta}
 + m\right]\psi^{\pm}_{{\bf k}s}(\tau) = 0,
\label{boost_mode_eq_D}\end{eqnarray}
Squaring the Dirac equation:
\begin{equation}
\psi^{\pm}_{{\bf k}s} = \left[-\gamma^{0}\left( {d \over d\tau}
+{1 \over 2\tau}\right)
- i \gamma_{\bf{{\perp}}}\cdot {\bf{k_{\perp}}}
-i \gamma^{3} \pi_{\eta}+ m\right] \chi_{s} {f^{\pm}_{{\bf k}s} \over
{\sqrt \tau}}
\, . \label{boost_psi_g}
\end{equation}
\begin{equation}
\gamma^{0}\gamma^{3}\chi_{s} = \lambda_{s} \chi_{s}\,
\end{equation}
with $\lambda_{s}=1$ for $s=1,2$ and $\lambda_{s}=-1$ for $s=3,4$,
we then get the mode equation:
\begin{equation}
\left( {d^2 \over d \tau^2}+
\omega_{\bf k}^2 -i \lambda_{s} \dot{\pi}_{\eta} \right )
f^{\pm}_{{\bf k}s}(\tau) = 0,
\label{boost_modef_D}
\end{equation}
\begin{equation}
 \omega_{\bf k}^2= \pi_{\eta}^2 +{\bf k}_{\perp} ^2 +m^2;~~
\pi_{\eta}={k_{\eta} -eA \over \tau}. \label{boost_omega_D}
\end{equation}

The back-reaction equation  in terms of the modes  is
\begin{equation}
{1 \over \tau}{dE(\tau) \over d\tau} = -{2e \over \tau^2}
\sum_{s=1}^{4}\int [d{\bf k}]
({\bf k}^2_{\perp} +m^2)
\lambda_{s}\vert f_{{\bf k}s}^{+}\vert ^2 ,
\label{boost_MaxD3}
\end{equation}
A typical proper time evolution of $E$ and $j$ is shown in
fig. 4. Here an initial value of $E=4$ was chosen, and here $E$ is the dimensionless
$E/E_{critical}$ relevant for unsuppressed pair production.

%
%
\subsection{Spectrum of particles}

To determine the number of particles produced one needs to introduce
the adiabatic bases for the fields:
\begin{eqnarray}
\Psi (x) &&= \int [d{\bf k}] \sum_{s}[b_{s}^{0}({\bf k};\tau)
u_{{\bf k}s}(\tau)
 e^{-i \int \omega_{\bf{k}} d \tau} \nonumber \\
&&+d_{s}^{(0) \dagger}({\bf{-k}};\tau) v_{{\bf -k}s}(\tau)
 e^{i \int \omega_{\bf{k}}d \tau}] e^{i {\bf k \cdot x}}.
\label{adiab}
\end{eqnarray}

The operators $b_s ({\bf k})$ and $b_{s}^{(0)}({\bf k};\tau)$ are
related by a Bogolyubov transformation:
\begin{eqnarray}
b_{r}^{(0)}({\bf k};\tau) &=& \sum \alpha_{{\bf k} r}^s(\tau) b_s({\bf
k}) + \beta_{{\bf k} r}^s(\tau) d_s^{\dag}({\bf -k}) \nonumber \\
d_{r}^{(0)}({\bf -k};\tau) &=& \sum \beta_{{\bf k} r}^{*s}(\tau)
b_s({\bf k}) + \alpha_{{\bf k} r}^{*s}(\tau) d_s^{\dag}({\bf -k})
\end{eqnarray}
One finds that the interpolating phase space number density for the
number of particles (or antiparticles) present per unit phase space
volume at time $\tau$ is given by:
\begin{equation}
n({\bf k};\tau) = \sum_{r=1,2} \langle 0_{in} |b_{r}^{(0) \dag}({\bf
k};\tau) b_{r}^{(0)}({\bf k};\tau) |0_{in} \rangle = \sum_{s,r}
|\beta^s_{{\bf k} r}(\tau) |^2
\end{equation}

This is an adiabatic invariant of the Hamiltonian dynamics governing
the time evolution of the one and two point functions, and is
therefore the logical choice as the interpolating particle number operator. At
$\tau=\tau_0$ it is equal to our initial number operator. If at later
times one reaches the out regime because of the decrease in energy
density due to expansion it becomes the usual out state phase space
number density.

The phase space distribution of particles (or antipartcles) in light
cone variables is
\begin{equation}
n_{\bf k}(\tau) =f(k_{\eta}, k_{\perp},\tau) = {d^6 N \over \pi^2
dx_{\perp}^2 dk_{\perp}^2d\eta dk_{\eta}}.
\label{interplc}
\end{equation}

We now need to relate this quantity to the spectra of electrons and
positrons produced by the strong electric field.  We introduce the particle rapidity $y$ and
$m_{\perp}= \sqrt{k_{\perp}^2 + m^2}$ defined by the particle
4-momentum in the center of mass coordinate system
\begin{equation}
k_{\mu} = (m_{\perp} \cosh y, k_{\perp},m_{\perp} \sinh y)
\label{kmu}
\end{equation}
The boost that takes one from the center of mass coordinates to the
comoving frame where the energy momentum tensor is diagonal is given
by $\tanh \eta= v = z/t$, so that one can define the ``fluid"
4-velocity in the center of mass frame as
\begin{equation}
u^{\mu} = (\cosh \eta,0,0, \sinh \eta)
\end{equation}
We then find that the variable
\begin{equation}
\omega_k = \sqrt{m_{\perp}^2+{ k_{\eta}^2 \over \tau^2} } \equiv
k^{\mu}u_{\mu}
\end{equation}
has the meaning of the energy of the particle in the comoving frame.
The momenta $k_{\eta}$ that enters into the adiabatic phase space
number density is one of two momenta canonical to the variables
defined by the coordinate transformation to light cone
variables. Namely the variables
\begin{eqnarray}
 \tau = (t^2-z^2)^{1/2}  \qquad \,
\eta = \frac{1}{2} \ln \left({{t+z} \over{t-z}} \right) \nonumber
\end{eqnarray}
have as their canonical momenta
\begin{eqnarray}
k_{\tau}= Et/ \tau -k_z z/ \tau \quad \, k_{\eta}  = -Ez + t k_z.
\label{boost_4trans}
\end{eqnarray}
To show this we consider the metric $ ds^2= d\tau^2 - \tau^2 d \eta^2$
and the free Lagrangian
\begin{equation}
L = {m \over 2} g_{\mu \nu} {dx^{\mu} \over ds} {dx^{\nu} \over ds}
\end{equation}
Then we obtain for example
\begin{eqnarray}
k_{\tau} &=& m {d \tau \over ds} = m [({\partial \tau \over \partial
t})_z {d t
\over ds}+ ({\partial \tau \over \partial z})_t {d z \over ds}]
\nonumber \\
&=& {Et -k_z z  \over \tau} = k^{\mu} u_{\mu}
\end{eqnarray}
The interpolating phase-space density $f$ of particles depends on
$k_{\eta}$, ${\bf k}_{\perp}$, $\tau$, and is $\eta$-independent.  In
order to obtain the physical particle rapidity and transverse momentum
distribution, we change variables from $(\eta, k_{\eta})$ to $(z, y)$
at a fixed $\tau$ where $y$ is the particle rapidity. We have
\begin{equation}
E{d^3 N \over d^3 k} =
{d^3N \over  \pi dy\,dk_{\perp}^2 } =
\int\pi dz~ dx_{\perp}^2 ~ J ~ f(k_{\eta},
k_{\perp},\tau)
 \label{boost_J}
\end{equation}
where the Jacobian $J$ is evaluated at a fixed proper time $\tau$
\begin{equation}
 J  = \left| \begin{matrix}
 {\partial k_{\eta}}/{\partial y}&{\partial k_{\eta}}/{\partial z} \cr
{\partial \eta}/{\partial y}& {\partial \eta}/{\partial z}  \end{matrix} \right|
\end{equation}

\begin{equation}  J= { m_{\perp} \cosh(\eta-y) \over
\cosh \eta}={\partial k_{\eta}\over \partial z}|_{\tau} .
\label{boost_Jinv}
\end{equation}
We also have
\begin{eqnarray}
k_{\tau} = m_{\perp} \cosh(\eta-y); \quad
k_{\eta} = -\tau m_{\perp} \sinh(\eta-y)\,.
\label{boost_ptaueta}
\end{eqnarray}
 Calling the integration over the transverse dimension the effective
transverse size of the colliding ions $A_{\perp}$ we then obtain that:
\begin{equation}
{d^3N \over  \pi dy\,dk_{\perp}^2} = A_{\perp} \int dk_{\eta}
f(k_{\eta}, k_{\perp},\tau) \equiv {d^3N \over  \pi d\eta\,dk_{\perp}^2}
\end{equation}
This quantity is independent of $y$ which is a consequence of the
assumed boost invariance. Note that we have proven using the property
of the Jacobean, that the distribution of particles in partical
rapidity is the same as the distribution of particles in fluid
rapidity verifying that in the boost-invariant regime that Landau's
intuition was correct  \cite{Landau}.

We now want to motivate the Cooper- Frye formula used to calculate particle spectrum in
hydrodynamical models of particle production \cite{cooper-frye}.  First we
note that the interpolating number density depends on $k_{\eta}$ and
$k_{\perp}$ only through the combination:
\begin{equation}
\omega_k = \sqrt{m_{\perp}^2+{ k_{\eta}^2 \over \tau^2} } \equiv
k^{\mu}u_{\mu}
\end{equation}
Thus $f(k_{\eta},k_{\perp}) = f(k_{\mu} u^{\mu})$ and so it depends on
exactly the same variable as the comoving thermal distribution! We
also have that a constant $\tau$ surface (which is the freeze out
surface of Landau) is parametrized as:
\begin{equation}
d\sigma^{\mu} = A_{\perp} (dz,0,0,dt)
= A_{\perp}d\eta (\cosh \eta ,0,0,\sinh \eta )
\end{equation}
We therefore find
\begin{equation}
k^{\mu} d\sigma_{\mu} =  A_{\perp} m_{\perp} \tau \cosh(\eta-y)=
A_{\perp} |
dk_{\eta} |
\end{equation}
Thus we can rewrite our expression for the field theory particle
spectra as
\begin{equation}
{d^3N \over  \pi dy\,dk_{\perp}^2} = A_{\perp} \int dk_{\eta}
f(k_{\eta}, k_{\perp},\tau)= \int f(k^{\mu}u_{\mu},\tau) k^{\mu}
d\sigma_{\mu}
\end{equation}
where in the second integration we keep $y$ and $\tau$ fixed.  Thus
with the replacement of the thermal single particle distribution by
the interpolating number operator, we get via the coordinate
transformation to the center of mass frame the Cooper-Frye formula.

\begin{figure}[t!]
   \center
   \includegraphics[width=6.0in]{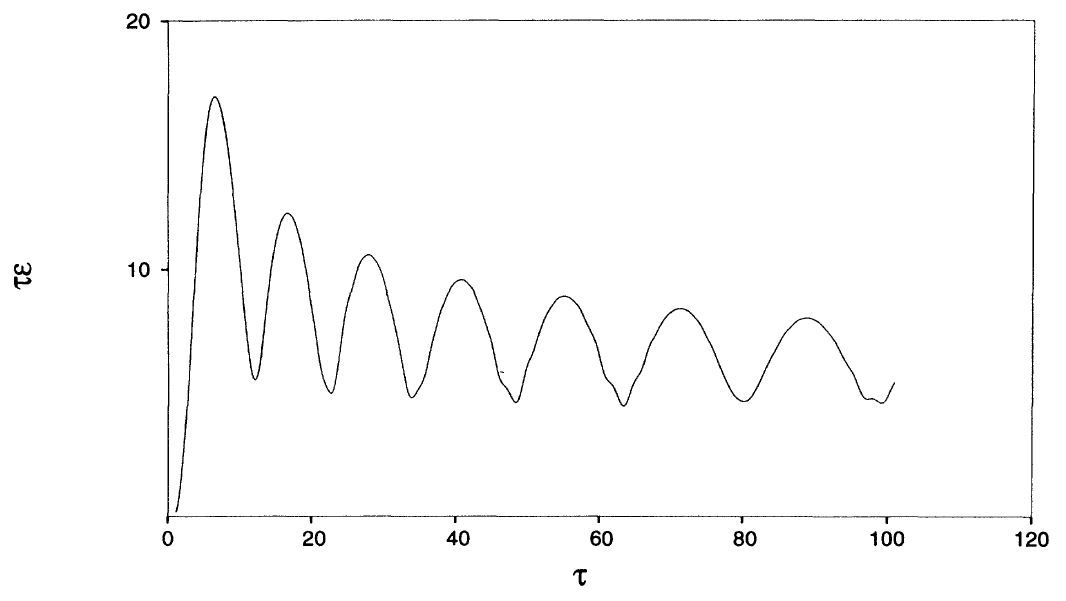}
   \caption{\label{f:plt7}Time evolution of $\varepsilon \tau$
   as a function of $\tau$.}
\end{figure}

\begin{figure}[t!]
   \center
   \includegraphics[width=6.0in]{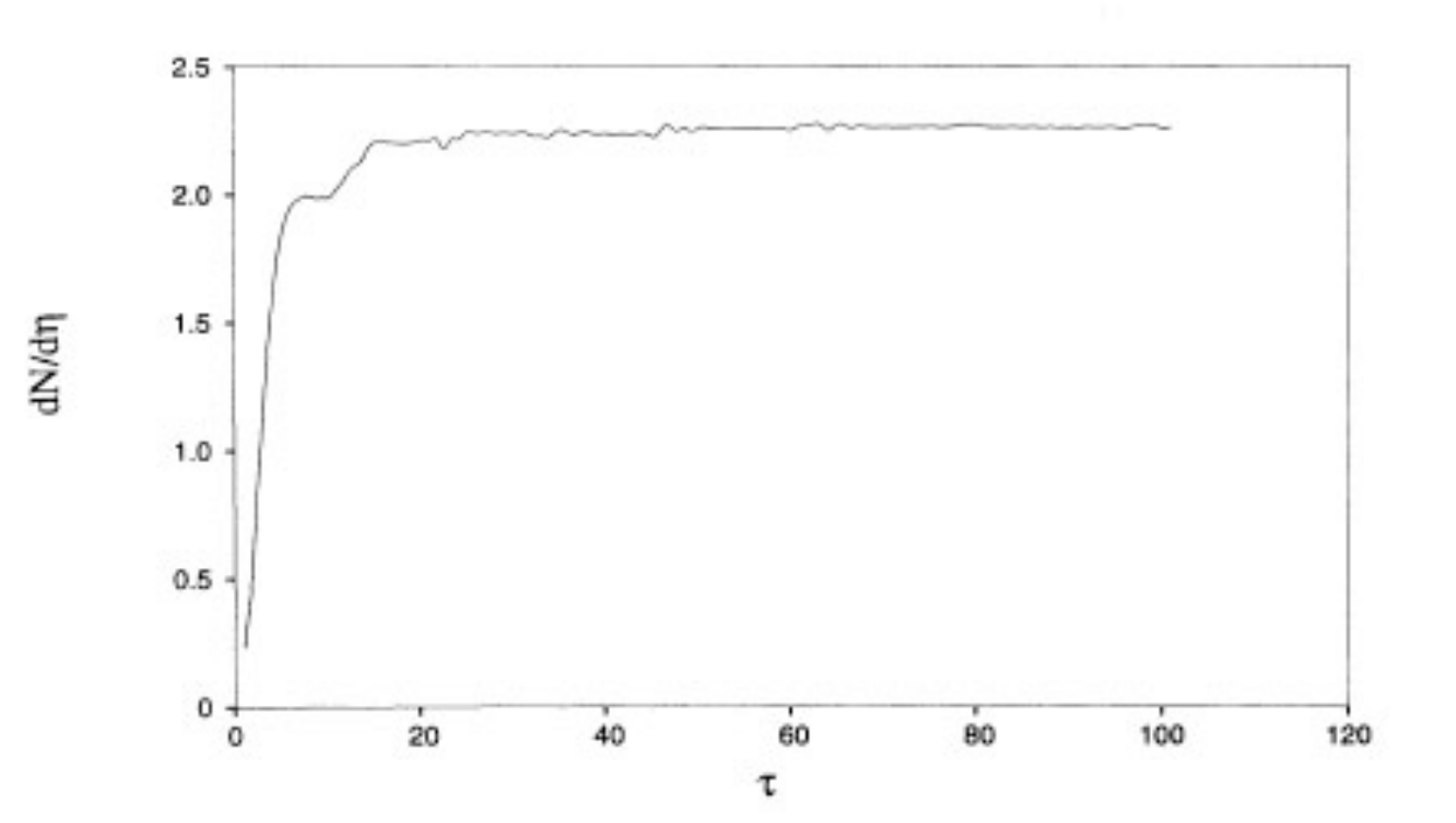}
   \caption{\label{f:plt8}The ratio of the approximate rapidity
   distribution $E / ( \,  m \cosh y \, ) \, ( d N / d y)$ and
   exact rapidity distribution as a function of $\tau$.}
\end{figure}

\begin{figure}[t!]
   \center
   \includegraphics[width=6.0in]{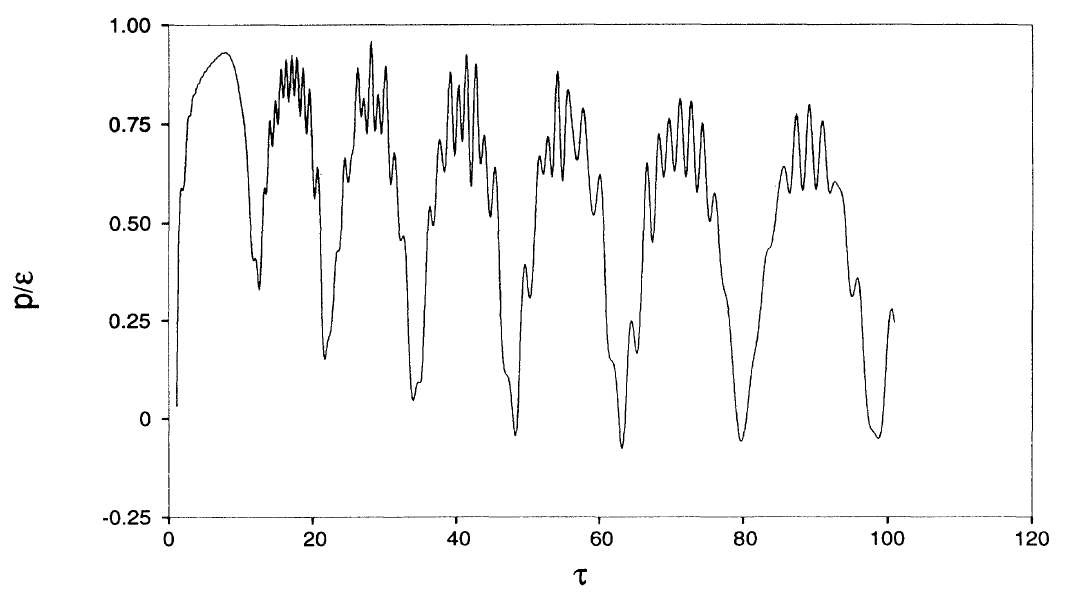}
   \caption{\label{f:plt9}Proper time evolution of $p / \varepsilon$.}
\end{figure}

The boost invariant assumption leads to an Energy Momentum
tensor which is diagonal in the($\tau,\eta, x_{\bot}$) coordinate
system which is thus a comoving one.  In that system one has:
\begin{equation}
{T^{\mu\nu} = {\rm diagonal}\; \lbrace \varepsilon(\tau),
p_{\parallel}(\tau), p_{\bot}(\tau), p_{\bot}(\tau) \rbrace}
\end{equation}
We thus find in this approximation that there are two separate
pressures, one in the longitudinal direction and one in the transverse
direction which is quite different from the thermal equilibrium
case. However only the longitudinal pressure enters into the
``entropy'' equation,
\begin{equation}
\varepsilon + p_{\parallel} = Ts \label{eq:entro1}
\end{equation}
\[ {d(\varepsilon \tau)\over d \tau} + p_{\parallel} = E j_{\eta} \]
\[ {d(s\tau) \over  d \tau}= { E j_{eta} \over T} \]
In the out regime we  find as in the Landau Model that $ s \tau = \text{constant}$.  The energy density as a function of proper time is shown in Fig.~\ref{f:plt7}.
%
%
%
%
For a one-dimensional boost invariant flow we find that the energy in a bin of fluid rapidity is just:
\begin{equation}
{dE \over d\eta} = \int T^{0\mu} d\sigma_{\mu} =
 A_{\bot} \tau \cosh \eta \varepsilon (\tau)
\end{equation}
which is just the $(1 + 1)$ dimensional hydrodynamical result.  Here
however $\varepsilon$ is obtained by solving the field theory equation
rather than using an ultrarelativistic equation of state.  This result
does not depend on any assumptions of thermalization.  We can ask if
we can directly calculate the particle rapidity distribution from the
ansatz that we divide the energy in a bin of fluid rapidity by the energy of a particle
assuming that the fluidity rapidity is equal to the particle rapidity?
\begin{equation}
{dN \over d \eta} = {1 \over m \cosh \eta} {dE \over d \eta} = { A_{\bot}
\over m} \varepsilon (\tau) \tau.
\end{equation}
We see from Fig.~\ref{f:plt8} that this works well even in our case where we have ignored rescattering, so that one does not have an equilibrium equation of state.
%
%
%
%
In the field theory calculation the expectation value of the stress
tensor must be renormalized since the electric field undergoes charge
renormalization.  We can determine the two pressures and the energy
density as a function of $\tau$.  Explicitly we have in the fermion
case.
\begin{equation*}
  \varepsilon (\tau)
   =
   \Expect{ T_{\tau\tau} }
   =
   \tau \Sigma_{s} \,
   \int [dk] R_{\tau\tau}(k) + E_{R}^{2}/2 \>,
\end{equation*}
where
\begin{eqnarray}
&& R_{\tau\tau}(k) =2(p_{\bot}^{2}
+ m^{2}) (g_{0}^{+}| f^{+}|^{2} - g_{0}^{-}| f^{-}|^{2}) - \omega
\nonumber \\
&&- (p_{\bot}^{2} + m^{2}) (\pi + e \dot{A})^{2}/( 8
\omega^{5} \tau^{2})\nonumber \\
&& p_{\parallel} (\tau) \tau^{2} = < T_{\eta \eta} > = \tau \Sigma_{s}
\int [dk]
\lambda_{s}\pi R_{\eta\eta}(k) - {1\over 2} E_{R}^{2} \tau^{2}
\end{eqnarray}
where
\begin{eqnarray}
R_{\eta\eta} (k) && =2 | f^{+}|^{2} -
(2\omega)^{-1}(\omega+\lambda_{s}\pi)^{-1}
 - \lambda_{s} e\dot{A}
/8\omega^{5}\tau^{2} \nonumber \\
&& -\lambda_{s} e\dot{E}/8\omega^{5}
-\lambda_{s}\pi/4\omega^{5}\tau^{2} + 5\pi\lambda_{s} (\pi + e
\dot{A})^{2}/(16 \omega^{7}\tau^{2}) \nonumber \\
\end{eqnarray}
and
\begin{eqnarray}
&& p_{\bot}(\tau)= < T_{yy} > = <T_{xx} > \nonumber \\
&& =(4 \tau)^{-1}
\sum_{s} \int [dk] \lbrace p_{\bot}^{2}(p_{\bot}^{2} +m^{2})^{-1}
R_{\tau\tau} -2\lambda\pi p_{\bot}^{2} R_{\eta\eta}\rbrace \nonumber \\
&&+ E_{R}^{2}/2.
\end{eqnarray}
Thus we are able to numerically determine the dynamical equation of
 state $p_{i}=p_{i}(\varepsilon)$ as a function of $\tau$.  A typical
 result is shown in Fig.~\ref{f:plt9}.

%
%
%
%

%
%

\section{QCD back-reaction problem with cylindrical symmetry}
\label{s:QCDcylindricalsym}

Previously we have discussed the constant Chromoelectric field problem and have shown that the answer for gluon production is independent of the Gauge Fixing term in covariant gauges. The answer depended on two Casimirs.  If the Schwinger mechhanism is important for these experiments in producing the initial plasma, one might be able to find an experimentally observable effect at RHIC and LHC for those jets coming from the semi-classical gluon plasma produced at RHIC following the collision of heavy Ions. One possibility is that event by event the transverse distribution of jets produced following a heavy ion collision might depend on the values of both Casimirs and not just the initial energy density present in the semi-classical gluonic field.
Here we will only consider pair production of  quarks, described by a field $\psi(x)$ and satisfying Dirac's equation:
\begin{equation}\label{T.e:DiracI}
   \bigl [ \,
      \gamma^{\mu} \,
      \bigl ( \, \partial_{\mu}
         -
         g \, A_{\mu}(x) \,
      \bigr )
      +
      m \,
   \bigr ] \, \psi(x) = 0 \>,
\end{equation}
interacting with a classical Yang-Mills field $A_{\mu}(x) = A^{a}_{\mu}(x) \, T^a$, where $T^a$ are the generators of the $SU(3)$ algebra, and satisfying a back-reaction equation given by:
\begin{equation}\label{e:backreactionI}
   D_{\mu}^{ab} \, F^{b,\mu\nu}(x) =
      g \,
   \langle \,
      \Comm{\hat{\bar{\psi}}(x)}
           {\tilde{\gamma}^{\nu}(x) \, T^a \, \hat{\psi}(x) } \,
   \rangle / 2 \>,
\end{equation}
with $D_{\mu}^{ab} = \delta^{ab}\partial_{\mu} + g \, f^{abc}\, A^c_{\mu}(x)$.

We want to consider cases when the plasma starts out in a hot equilibrium state with cylindrical symmetry, and then expands into the vacuum.  We introduce the  fluid rapidity ($\eta$) and proper
time $\tau$  coordinates which are components of  $x^{\mu}=(\tau,\rho,\theta,\eta)$, and which are related to Cartesian coordinates by:
\begin{equation}
   t= \tau \cosh \eta \>,
   \qquad
   z= \tau \sinh \eta \>,
   \qquad
   x= \rho \cos \theta \>,
   \qquad
   y= \rho \sin \theta \>.
\end{equation}
For a boost invariant expansion, the classical gauge fields are restricted to be in the $\eta$-direction and depend only on $\tau$.  We also consider only the $a=3$ and $a=8$ gauge fields which carry all colors.  This will allow us to explore the entire Casimir space.   Then, using the Gell-Mann representation for the $\lambda^a$ matrices,
\begin{equation}
\begin{split}
   \tilde{\gamma}^{\mu}(x) A_{\mu}(x) =
   &   \frac{1}{2} \, \tilde{\gamma}^{\eta}(x) \,
   \bigl [ \,
      A^3_{\eta}(\tau) \, \lambda^3
      +
      A^8_{\eta}(\tau) \, \lambda^8 \,
   \bigr ]
   \\ =
   &   \frac{1}{2} \, \tilde{\gamma}^{\eta}(x) \,
   \begin{pmatrix}
      A^3_{\eta}(\tau) + A^8_{\eta}(\tau)/\sqrt{3} & 0 & 0 \\
      0 & - A^3_{\eta}(\tau) + A^8_{\eta}(\tau)/\sqrt{3} & 0 \\
      0 & 0 & - 2 \, A^8_{\eta}(\tau)/\sqrt{3}
   \end{pmatrix} \>,
\end{split}
\end{equation}
The two Casimir invariants for $SU(3)$ are given by:
\begin{equation}
   C_1 = E^a E^a \>,
   \quad \text{and} \quad
   C_2=
      \bigl [ \,
      d^{abc} \, E^a \, E^b \, E^c \,
   \bigr ]^2 \>,
\end{equation}
where $d^{abc}$ are the symmetric $SU(3)$ structure factors.  Notice that $F^{a}_{\mu\nu} \, F^{a;\mu\nu}$ is a Lorentz invariant and equal to $C_1$.  In our coordinate system, Eq.~\eqref{e:backreactionI} becomes:
\begin{equation}\label{T.e:backreacttionII}
   - \frac{1}{\tau} \frac{\partial E^a}{\partial\tau}
   =
   \frac{g}{2} \,
   \langle \,
      \Comm{\hat{\bar{\psi}}(x)}
           {\tilde{\gamma}^{\eta}(x) \, T^a \, \hat{\psi}(x) } \,
   \rangle \>.
\end{equation}
Eqs.~\eqref{T.e:DiracI} and \eqref{T.e:backreacttionII} are the equations we want to solve.

For the fermi field operator, we put:
\begin{equation}\label{e:ferimfieldopI}
   \hat{\psi}(x)
   =
   S(\theta,\eta) \, \hat{\phi}(x) / \sqrt{\tau} \>,
\end{equation}
where $S(\theta,\eta) = S_{\rho}(\theta) \, S_{\tau}(\eta)$ is a Lorentz transformation given by:
\begin{equation}\label{e:LorentzS}
   S_{\rho}(\theta)
   =
   \Exp{ \theta \, \gamma^1 \gamma^2 / 2 } \>,
   \qquad
   S_{\tau}(\eta)
   =
   \Exp{ \eta \, \gamma^0 \gamma^3 / 2 } \>,
\end{equation}
and where $\hat{\phi}(x)$ now satisfies:
\begin{equation}\label{e:fermifieldeqI}
   \bigl [ \,
      \bar{\gamma}^{\mu}(\tau,\rho) \, \nabla_{\mu}
      -
      m \,
   \bigr ] \, \hat{\phi}(x) / \sqrt{\tau}
   =
   0 \>.
\end{equation}
Here $\bar{\gamma}^{\mu}(\tau,\rho)$ are given by:
\begin{equation}\label{e:bargammas}
   \bar{\gamma}^{\tau}
   =
   \gamma^0 \>,
   \qquad
   \bar{\gamma}^{\rho}
   =
   \gamma^1 \>,
   \qquad
   \bar{\gamma}^{\theta}(\rho)
   =
   \gamma^2 / \rho \>,
   \qquad
   \bar{\gamma}^{\eta}(\tau)
   =
   \gamma^3 / \tau \>,
\end{equation}
and $\nabla_{\mu} = \partial_{\mu} + \Pi_{\mu} + A_{\mu}(\tau)$ is the covariant derivative in the boost-invariant system:
\begin{equation}\label{e:Pidefs}
   \Pi_{\theta}
   =
   \gamma^1 \gamma^2 / 2 \>,
   \qquad
   \Pi_{\eta}
   =
   \gamma^0 \gamma^3 / 2 \>.
\end{equation}
In cylindrical coordinates the canonical fields obey:
\begin{equation}\label{e:anticommrotII}
   \AntiComm{\hat{\phi}_{\alpha}(\tau,\rho,\theta,\eta)}
            {\hat{\phi}_{\alpha'}^{\dagger}(\tau,\rho',\theta',\eta')}
   =
   \delta_{\alpha,\alpha'} \,
   \frac{\delta(\rho - \rho')}{\sqrt{ \rho \rho' }} \,
   \delta(\theta - \theta') \,
   \delta(\eta - \eta') \>,
\end{equation}
which is just the usual anticommutation relation.  We can now write the Dirac field operator in this curvilinear coordinate system by the expansion:
\begin{multline}\label{e:gensolII}
   \hat{\phi}(\tau,\rho,\theta,\eta)
   =
   \int_{-\infty}^{\infty} \! \frac{\rd \keta}{2\pi}
   \int_{0}^{\infty} \frac{ \kperp\, \rd \kperp }{ 2\pi }
   \sum_{h=\pm 1} \sum_{m=-\infty}^{+\infty}
   \\ \times
   \bigl \{ \,
      \hat{b}_{\keta,\kperp,m}^{(h)} \,
      \phi_{\keta,\kperp,m}^{(h,+)}(\tau,\rho,\theta,\eta)
      +
      \hat{d}_{\keta,\kperp,m}^{(h)\,\dagger} \,
      \phi_{-\keta,\kperp,-m}^{(-h,-)}(\tau,\rho,\theta,\eta) \,
   \bigr \} \>.
\end{multline}
where:
\begin{equation}\label{e:K1solutions}
   \phi_{k_{\perp},m}^{(h)}(\tau,\rho,\theta,\eta)
   =
   e^{i \keta \eta } \,
   \begin{pmatrix}
      \phi_{(+);\kperp}^{(h)}(\tau) \,
      \chi^{(h)}_{\kperp,m}(\rho,\theta)
      \\[6pt]
      \phi_{(-);\kperp}^{(h)}(\tau) \,
      \chi^{(-h)}_{\kperp,m}(\rho,\theta)
   \end{pmatrix} \>,
\end{equation}
with $\lambda = h k_{\perp}$, and where $h = \pm 1$.
\begin{equation}\label{e:chimat}
   \chi_{\kperp,m}^{(h)}(\rho,\theta)
   =
   \frac{1}{\sqrt{2}}
   \begin{pmatrix}
      e^{i m \theta} J_{m}(\kperp \rho) \\
      h \, e^{i (m+1) \theta} J_{m+1}(\kperp \rho)
   \end{pmatrix} \>,
\end{equation}
with eigenvalues $\lambda = h \kperp$ and the helicity $h = \pm 1$.  Orthogonality is given by the relation:
\begin{equation}\label{e:chiorthog}
   \int_{0}^{+\infty} \!\!\! \rho \, \rd \rho
   \int_{0}^{2\pi} \!\!\! \rd \theta \;
   \chi_{\kperp,m}^{(h)\dagger}(\rho,\theta) \,
   \chi_{\kperp',m'}^{(h')}(\rho,\theta)
   =
   \delta_{h,h'} \, \delta_{m,m'} \,
   ( 2\pi ) \,
   \frac{\delta(\kperp - \kperp')}
        {\sqrt{ \kperp \kperp' } }\>.
\end{equation}
The mode functions $\phi_{(\pm);\kperp}^{(h)}(\tau)$ satisfy:
\begin{equation}\label{e:tauequ}
   \begin{pmatrix}
      i \partial_{\tau} + m &
      - \pi_{\keta}(\tau) - i h \kperp
      \\
      - \pi_{\keta}(\tau) + i h \kperp
      &
      i \partial_{\tau} - m
   \end{pmatrix}
   \begin{pmatrix}
      \phi_{(+);\keta,\kperp}^{(h)}(\tau) \\
      \phi_{(-);\keta,\kperp}^{(h)}(\tau)
   \end{pmatrix}
   =
   0 \>,
\end{equation}
where we have defined $\pi_{\keta}(\tau)$ by:
\begin{equation}\label{e:pidef}
   \pi_{\keta}(\tau)
   =
   ( \, \keta - g \, A(\tau) \, ) / \tau \>.
\end{equation}
Eq.~\eqref{e:tauequ} is the equation we want to solve numerically as a function of $\tau$ for some given initial spinor at $\tau = 1$.

Now at $\tau = 1$, we will set $A(1) = 0$.  This means that \emph{near} $\tau = 1$, Eq.~\eqref{e:tauequ} becomes:
\begin{equation}\label{e:tauequatone}
   \begin{pmatrix}
      i \partial_{\tau} + m &
      - \keta - i h \kperp
      \\
      - \keta + i h \kperp
      &
      i \partial_{\tau} - m
   \end{pmatrix}
   \begin{pmatrix}
      \phi_{0\,(+);\keta,\kperp}^{(h)}(\tau) \\
      \phi_{0\,(-);\keta,\kperp}^{(h)}(\tau)
   \end{pmatrix}
   =
   0 \>,
\end{equation}
which have positive and negative frequency solutions of the form:
\begin{subequations}\label{e:psizeroposneg}
\begin{align}
   \phi_{0\,;\keta,\kperp}^{(h,+)}(\tau)
   &=
   \sqrt{ \frac{ \omega_{0;\keta,\kperp} - m }
               { 2 \omega_{0;\keta,\kperp} } } \,
   \begin{pmatrix}
      1
      \\
      \displaystyle
      +
      \frac{ \keta - i h \kperp }
           { \omega_{0;\keta,\kperp} - m }
   \end{pmatrix}
   \Exp { - i \omega_{0;\keta,\kperp} ( \tau - 1 ) }  \>,
   \label{e:psizeropos }\\
   \phi_{0\,;\keta,\kperp}^{(h,-)}(\tau)
   &=
   \sqrt{ \frac{ \omega_{0;\keta,\kperp} - m }
               { 2 \omega_{0;\keta,\kperp} } } \,
   \begin{pmatrix}
      \displaystyle
      -
      \frac{ \keta - i h \kperp }
           { \omega_{0;\keta,\kperp} - m }
      \\
      1
   \end{pmatrix}
   \Exp { + i \omega_{0;\keta,\kperp} ( \tau - 1 ) }  \>,
   \label{e:psizeroneg}
\end{align}
\end{subequations}
where $\omega_{0;\keta,\kperp} = \sqrt{ \keta^2 + \kperp^2 + m^2}$.
These solutions are orthogonal:
\begin{equation}\label{e:psizeroorthog}
   \sum_{\alpha = \pm}
   \phi_{0\,(\alpha);\keta,\kperp}^{(h,\lambda)\,\ast}(\tau) \,
   \phi_{0\,(\alpha);\keta,\kperp}^{(h,\lambda')}(\tau)
   =
   \delta_{\lambda,\lambda'} \>,
\end{equation}
and complete:
\begin{equation}\label{e:psizerocomplete}
   \sum_{\lambda = \pm 1}
   \phi_{0\,(\alpha);\keta,\kperp}^{(h,\lambda)}(\tau) \,
   \phi_{0\,(\beta);\keta,\kperp}^{(h,\lambda)\,\ast}(\tau)
   =
   \delta_{\alpha,\beta} \>.
\end{equation}
So at $\tau = 1$, we choose our solutions of Eq.~\eqref{e:tauequ} so that:
\begin{equation}\label{e:psistart}
   \phi_{(\alpha);\keta,\kperp}^{(h,\lambda)}(1)
   =
   \phi_{0\,(\alpha);\keta,\kperp}^{(h,\lambda)}(1) \>,
\end{equation}
for $\alpha = \pm$ and where $\lambda = \pm 1$ labels the initial positive and negative frequency solutions of Eq.~\eqref{e:tauequatone}.  The $\tau$-dependent solutions will then be numerically stepped out from the values at $\tau = 1$.

Maxwell's equation becomes:
\begin{equation}  \label{br.e:MaxwelltauIII}
   \partial_{\tau} E(\tau)
   = -   \frac{g}{\tau} \,
   \int_{-\infty}^{\infty} \! \frac{\rd \keta}{2\pi}
   \int_{0}^{\infty} \frac{ \kperp\, \rd \kperp }{ 2\pi }
   \sum_{h=\pm 1} \, j_{\keta,\kperp}^{(h)}(\tau) \>,
\end{equation}
where $j_{\keta,\kperp}^{(h)}(\tau)$ is given by the positive energy solutions of the Dirac equation only:
\begin{equation}\label{br.e:jdef}
\begin{split}
   j_{\keta,\kperp}^{(h)}(\tau)
   &=
   \phi_{(+);\keta,\kperp}^{(h,+)\,\ast}(\tau) \,
   \phi_{(-);\keta,\kperp}^{(h,+)}(\tau)
   +
   \phi_{(-);\keta,\kperp}^{(h,+)\,\ast}(\tau) \,
   \phi_{(+);\keta,\kperp}^{(h,+)}(\tau) \>,
   \\
   &=
   \phi_{\keta,\kperp}^{(h,+)\,\dagger}(\tau) \,
   \sigma_x \,
   \phi_{\keta,\kperp}^{(h,+)}(\tau) \>.
\end{split}
\end{equation}
Here, $\phi_{\keta,\kperp}^{(h,+)}(\tau)$ is the two-component positive energy spinor:
\begin{equation}\label{br.e:phitwocomponent}
   \phi_{\keta,\kperp}^{(h,+)}(\tau)
   =
   \begin{pmatrix}
      \phi_{(+);\keta,\kperp}^{(h,+)}(\tau)
      \\
      \phi_{(-);\keta,\kperp}^{(h,+)}(\tau)
   \end{pmatrix} \>,
\end{equation}
and $\sigma_x$ the Pauli matrix.  Dirac's Eq.~\eqref{e:tauequ} and Maxwell's
Eq.~\eqref{br.e:MaxwelltauIII}, are the update equations we want to solve simultaneously.

%
%

The definition of interpolating particle number for the time evolution problem is somewhat arbitrary. The main need is for it to interpolate between initial an final definitions of the number operator and for it to have some adiabatic properties.  We choose to define them here in terms of the exact solutions of the Dirac equation in the {\em absence} of external fields.  These zeroth order spinors are given by:
\begin{equation}\label{e.pp:modefcts}
   \phi_{0;\keta,\kperp,m}^{(h,\lambda)}(\tau,\rho,\theta,\eta)
   =
   e^{i \keta \eta} \,
   \begin{pmatrix}
      \phi_{0\,(+);\keta,\kperp}^{(h,\lambda)}(\tau) \,
      \chi^{(h)}_{\kperp,m}(\rho,\theta)
      \\[6pt]
      \phi_{0\,(-);\keta,\kperp}^{(h,\lambda)}(\tau) \,
      \chi^{(-h)}_{\kperp,m}(\rho,\theta)
   \end{pmatrix} \>,
\end{equation}
where $\phi_{0;\keta,\kperp}^{(h,\lambda)}(\tau)$ given by Eqs.~\eqref{e:psizeroposneg}.  These spinors are also orthogonal and complete.  Expansion of the field operator in the \emph{zeroth order} spinors then requires that the creation and annihilation operators $\hat{A}_{0;\keta,\kperp,m}^{(h,\lambda)}(\tau)$ become time-\emph{dependent}.  That is:
\begin{equation}\label{e:freesol}
   \hat{\phi}(\tau,\rho,\theta,\eta)
   =
   \int_{-\infty}^{\infty} \! \frac{\rd \keta}{2\pi}
   \int_{0}^{\infty} \frac{ \kperp\, \rd \kperp }{ 2\pi }
   \sum_{h=\pm 1}  \sum_{\lambda=\pm 1} \sum_{m=-\infty}^{+\infty}
   \hat{A}_{0;\keta,\kperp,m}^{(h,\lambda)}(\tau) \,
   \phi_{0;\keta,\kperp,m}^{(h,\lambda)}
      (\tau,\rho,\theta,\eta) \>,
\end{equation}
Because of the orthogonality of the initial spinors, we see that the $\hat{A}_{0;\keta,\kperp,m}^{(h,\lambda)}(\tau)$ operators obey the same commutation relations as the time-\emph{independent} ones at equal time:
\begin{equation}\label{e:A0A0anticomm}
   \AntiComm{\hat{A}_{0;\keta,\kperp,m}^{(h,\lambda)}(\tau)}
      {\hat{A}_{0;\keta',\kperp',m'}^{(h',\lambda')\,\dagger}(\tau)}
   =
   \delta_{\lambda,\lambda'} \,
   \delta_{h,h'} \,
   \delta_{m,m'} \,
   ( 2\pi )^2 \, \delta(\keta - \keta') \,
   \frac{\delta(\kperp - \kperp')}
        {\sqrt{ \kperp \kperp' }} \>,
\end{equation}
and so we can use them to define number operators at time $\tau$.  Again,
it is traditional to define particle and anti-particle operators at time $\tau$ by:
\begin{equation}\label{e:A0opdefs}
   \hat{A}_{0;\keta,\kperp,m}^{(h,+)}(\tau)
   =
   \hat{b}_{0;\keta,\kperp,m}^{(h)}(\tau) \>,
   \qquad\text{and}\qquad
   \hat{A}_{0;\keta,\kperp,m}^{(h,-)}(\tau)
   =
   \hat{d}_{0;-\keta,\kperp,-m}^{(-h)\,\dagger}(\tau) \>.
\end{equation}
As before we can determine the adiabatic number operator from the Bogoliubov transformation.
The overlap between the adiabatic wave functions and the exact ones is :
 $C_{\keta,\kperp}^{(h;\lambda,\lambda')}(\tau)$ is given by:
\begin{equation}\label{e:Cdef}
   C_{\keta,\kperp}^{(h;\lambda,\lambda')}(\tau)
   =
   \phi_{0;\keta,\kperp}^{(h,\lambda)\,\dagger}(\tau) \,
   \phi_{\keta,\kperp}^{(h,\lambda')}(\tau) \>,
\end{equation}
and is independent of $m$.  So the creation and annihilation operators are related by the expression:
\begin{equation}\label{e:A0toA}
   \hat{A}_{0;\keta,\kperp,m}^{(h,\lambda)}(\tau)
   =
   \sum_{\lambda = \pm}
   C_{\keta,\kperp}^{(h;\lambda,\lambda')}(\tau) \,
   \hat{A}_{\keta,\kperp,m}^{(h,\lambda')} \>,
\end{equation}
which is a Bogoliubov transformation of the operators.
Now let $n_{\kperp,\phi,\keta}^{(h)}(\tau)$ be the number of particles  produced at time $\tau$ per unit volume in a given mode and helicity $h$:
\begin{equation}\label{if.e:ndef}
   n_{\keta,\kperp,\phi}^{(h)}(\tau)
   =
   \frac{\rd^6 N (\tau) }
      { \rd \keta \, \kperp \, \rd \kperp \, \rd \phi \,
        \rd \eta \, \rho \, \rd \rho \, \rd \theta } \>.
\end{equation}
$n_{\keta,\kperp,\phi}^{(h)}(\tau)$ is given by the formula:
\begin{equation}\label{if.e:nvalue}
   \Expect{
      \hat{a}_{0\,\keta,\kperp,\phi}^{(h)\dagger}(\tau) \,
      \hat{a}_{0\,\keta',\kperp',\phi'}^{(h')}(\tau)
          }
    =
    n_{\keta,\kperp,\phi}^{(h)}(\tau) \,
    \delta_{h,h'} \,
    ( 2\pi )^3 \,
    \delta( \keta - \keta' ) \,
    \frac{ \delta( \kperp - \kperp' ) }
         { \sqrt{ \kperp \kperp' } } \,
    \delta( \phi - \phi' ) \>,
\end{equation}

\begin{equation}\label{e:nvalueC}
   n_{\keta,\kperp,\phi}^{(h)}(\tau)
   =
   \frac{\rd^6 N (\tau) }
      { \rd \keta \, \kperp \, \rd \kperp \, \rd \phi \,
        \rd \eta \, \rho \, \rd \rho \, \rd \theta }
   =
   | \, C_{\keta,\kperp}^{(h;+,-)}(\tau) \, |^2
   =
   1
   -
   | \, C_{\keta,\kperp}^{(h;+,+)}(\tau) \, |^2 \>,
\end{equation}
and is \emph{independent} of $\phi$.    Explicitly, $| \, C_{\keta,\kperp}^{(h;+,+)}(\tau) \, |^2$ is given by
\begin{equation}\label{e:C2calc}
   | \, C_{\keta,\kperp}^{(h;+,+)}(\tau) \, |^2
   =
   \frac{ \omega_{\keta,\kperp} - m }
        { 2 \omega_{\keta,\kperp}   } \,
   \Big | \,
      \phi_{(+);\keta,\kperp}^{(h,+)}(\tau)
      +
      \frac{ \keta + i h \, \kperp }
           { \omega_{\keta,\kperp} - m  } \,
      \phi_{(-);\keta,\kperp}^{(h,+)}(\tau) \,
   \Big |^2 \>.
\end{equation}
The current vanishes at $\tau=1$, as required.

%
%
\section{Flavor large-N expansion in a 2PI approach}

It is important to know whether rescatterings due to interactions betweeen the quarks and gluons will change the final transverse distribution functions from those found in the leading order in
flavor large-N. To include interactions among the quarks and gluons one would solve the coupled Schwinger-Dyson equations using the CTP formalism and a 2PI \cite{2PI} effective action expanded in 1/N  \cite{noneq}.  Here one would need to keep the background field formalism also to handle the background chromoelectric field.  This formalism has been used in QCD to determine transport coefficients by Aarts and Resco \cite{Aarts}.  The action for $N_f$ identical fermion fields $\psi_a$ ($a=1,\ldots,N_f$) then reads (We use here  $g_{\mu\nu}=\mbox{diag}(+,-,-,-)$)

\beq
 S = \int_x \left[ -\frac{1}{4}F_{\mu\nu}F^{\mu\nu}
 +\bar\psi_a \left(i\Dslash - m \right)\psi_a \right]
 + S_{\rm gf} +S_{\rm gh},
\ee
with
\beq
 \label{eqcov}
 \Dslash = \gamma^\mu D_\mu, \;\;\;\;\;\;\;\;
 D_\mu = \partial_\mu + \frac{ie}{\sqrt{N_f}} A_\mu,
\ee
and we use the notation
\beq
 \int_x = \int_{\cal{C}} dx^0 \int d^3 x,
\ee
where ${\cal{C}}$ refers to the CTP  contour in the complex-time plane. Note that we have rescaled the coupling constant with $\sqrt{N_f}$, so that in the large $N_f$ limit $N_f$ goes to infinity while $e$ remains finite after renormalization.  To fix the gauge one can use a general linear gauge fixing condition,
\beq
 S_{\rm gf} = -\int_x \frac{1}{2\xi} \;
    [ \, f_\mu A^\mu \, ]^2 \>.
\ee
Below we specialize to the generalized coulomb gauge: $f_0=0,
f_i=\partial_i$. The ghost part is not needed explicitly.

\begin{figure}[t!]
   \center
   \includegraphics[width=0.8in]{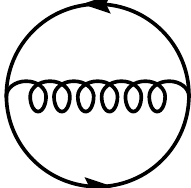}
   \caption{\label{f:plt10}NLO contribution to the 2PI effective
   action in the $1/N_f$ expansion.}
\end{figure}

The 2PI effective action is an effective action for the contour-ordered two-point functions
\beq
   D_{\mu\nu}(x,y)
   =
   \langle \,
   \mathcal{T}_\mathcal{C} \,
   \bigl [ \, A_\mu(x) \, A_\nu(y) \, \bigr ] \, \rangle \>,
   \qquad
   S_{ab}(x,y)
   =
   \langle \,
   \mathcal{T}_\mathcal{C} \,
   \bigl [ \, \psi_a(x)\bar\psi_b(y) \, \bigr ] \,
   \rangle
\ee
and can be written as
\begin{multline}
   \Gamma[S, D]
   =
   \frac{i}{2} \,
   \Tr{ \Ln{ D^{-1} } }
   +
   \frac{i}{2} \,
   \Tr{ D_0^{-1}(D-D_0) }
   \\
   - i \,
   \Tr{ \Ln{ S^{-1} } }
   -
   i \,
   \Tr{ S_0^{-1}(S-S_0) }
   +
   \Gamma_2[S,D]
   +
   \text{ghosts} \>,
\end{multline}
where $D_0^{-1}$ and $S_0^{-1}$ are the free inverse propagators and $\Gamma_2$ contains all two particle irreducible vacuum graphs such as those in Fig.~\ref{f:plt10}.

%
%
\begin{figure}[t!]
   \center
   \includegraphics[width=2.5in]{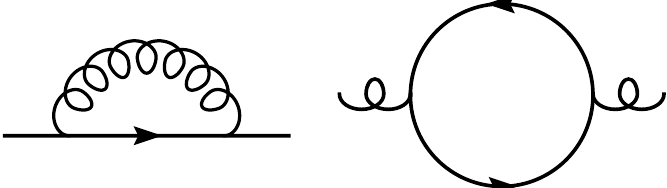}
   \caption{\label{f:plt11}Self energies at next-to-leading order
   in large $N_f$ QCD.}
\end{figure}

%
%
For the gauge
theory the next-to-leading order Schwinger-Dyson equations that result from varying the 2PI action are:
\begin{equation*}
   S^{-1} = S_0^{-1} - \Sigma \>,
   \qquad
   D^{-1} = D_0^{-1} - \Pi \>,
\end{equation*}
with $S$ the fermion and $D$ the gauge field propagator.  The self
energies, depending on full propagators, are shown in Fig.~\ref{f:plt11}.
%
%

%
%
%
%
The back reaction equation is given by:
\beq
\nabla_\mu F^{\mu \nu} = \langle j^\nu \rangle = -i g^2 Tr{\gamma^\nu S}
\ee
To determine the interpolating number densities of quarks and antiquarks one can follow the procedure of Berges, Borsanyi and Serreau \cite{BBS} and define these from the current.  Namely the associated $4$-current for each given flavor is
$\sim \bar\psi \gamma^\mu \psi$.  Fourier transforming with respect to spatial momenta, the expectation value of
the latter can be written as
$J_f^\mu (t,p) = \Tr{ \gamma^\mu S^<(t,t,p) }$,
In terms of the equal-time two point function, its temporal and spatial components are
\begin{align*}
   J_f^0 (t,p)
   &=
   2 \, [ \, 1 - 2\,F_V^0(t,t;p) \, ] \>,
   \\
   \mathbf{J}_f (t,p)
   &=
   - 4 \, F_V(t,t;p) \>.
\end{align*}
To obtain an effective particle number, Berges et. al.
 identify these expressions with the corresponding ones in a
quasi-particle description with free-field expressions. These
are given by
\begin{align*}
   J_{f}^{0\, (\text{QP})}(t,p)
   &=
   2 \, [1+Q_f(t,p)] \>,
   \\
   \mathbf{J}_{f}^{\,(\text{QP})}(t,p)
   &=
   - 2  \,[ \, 1- 2 N_f(t,p) \, ] \>,
\end{align*}
where $Q_f(t,p)=n_f-\bar{n}_f$ is the difference between particle and
antiparticle effective number densities and $N_f(t,p)=(n_f+\bar{n}_f)/2$ is their half-sum. The physical content of these expressions is simple: the temporal component $J^0$ directly represents the net charge density per mode $Q_f(t,p)$, whereas the spatial part $\mathbf{J}$ is the net current density per mode and is therefore sensitive to the sum of particle and antiparticle number densities.  Identifying the above expressions, they define
\begin{align*}
   \frac{1}{2}\,Q_f(t,p)
   &=
   - F_V^0(t,t;p) \>,
   \\
   \frac{1}{2} - N_f(t,p)
   &=
   F_V(t,t;p) \>.
\end{align*}
Using these definitions and solving the backreaction problem to NLO in 2PI-1/N we would also be able to discover if there is time for the produced quarks and antiquarks to thermalize  before hadronization time scale and to see if the constant field result for the transverse distribution will be modified by the interactions.  This will be an important calculation to do in the future.

%
%
\acknowledgments

We would like to thank all our collaborators, Emil Mottola, So-Young Pi, Yuval Kluger, Salman Habib, Juan Pablo Paz, and Gouranga Nayak for sharing their ideas, enthusiasm and efforts during this project.  This work was supported in part by the Department of Energy and by National Science Foundation, grants PHY-0354776 and PHY-0345822.  Fred Cooper would like to thank Harvard University for providing hospitality.  All of us would like to thank the Santa Fe Institute for hospitality at various times during this research.

%
%

\end{document}